\documentclass[prl,aps,twocolumn,floats,showpacs]{revtex4}
\usepackage{bm,graphicx}

\newcommand{\be}{\begin{equation}}
\newcommand{\ee}{\end{equation}}
\newcommand{\bea}{\begin{eqnarray}}
\newcommand{\eea}{\end{eqnarray}}
\newcommand{\phrl}[1]{Phys.~Rev.~Lett. {\bf #1}}
\newcommand{\phrb}[1]{Phys.~Rev.~B {\bf #1}}
\newcommand{\cmat}[1]{arXiv:cond-mat/#1}
\newcommand{\bib}{\bibitem}
\newcommand{\lb}{\left[}
\newcommand{\rb}{\right]}
\newcommand{\lp}{\left(}
\newcommand{\rp}{\right)}
\newcommand{\lf}{\left\{}
\newcommand{\rf}{\right\}}

\newcommand{\G}{{\cal G}}
\renewcommand{\r}{{\bf r}}
\renewcommand{\j}{{\bf j}}
\newcommand{\p}{{\bf p}}
\newcommand{\q}{{\bf q}}
\renewcommand{\k}{{\bf k}}
\newcommand{\J}{{\bf J}}

\begin{document}

\title{`Spin-Spin' Hall Effect in Two-Dimensional
         Electron Systems with Spin-Orbit Interaction}
\author{Ankur Sensharma and Sudhansu S. Mandal}
\affiliation{Theoretical Physics Department, Indian Association for the 
   Cultivation of Science, Jadavpur, Kolkata 700 032, India}

\date{\today}


\begin{abstract}
We calculate spin-Hall conductivities
in the two dimensional electron systems with Rashba spin-orbit interaction.
The salient feature is that, apart from the
usual spin-Hall conductivity $\sigma^z_{xy}$ which corresponds to the 
induction of out-of-plane
spin-current due to the application of transverse charge current, 
there is a novel spin-Hall conductivity $\sigma^\perp_{xy}$ which arises
due to the induction of transverse spin-polarized current in the transverse 
direction by the application of in-plane spin-polarized current.
This phenomenon which we call as `spin-spin' Hall effect is a spin
analog of conventional Hall effect, but with no magnetic field.
This contribution may be understood through the 
spin-diffusive equation. 

\end{abstract}

\maketitle


\section{Introduction} 

Following pioneering proposal of a device called spin field effect
transistor (SFET) by 
Datta and Das\cite{Datta}, there has been growing interest in the field of 
spintronics\cite{Sarma} which is the
science of coherent manipulation of spin,
because of its potential applications in spin-memory
and quantum computing devices\cite{Review}.
This device consists of
a two-dimensional electron system (2DES), controlled by gate voltage
in a semiconductor heterostructure with spin polarized contacts.
A spin entered from a spin-polarized source in the 2DES, precesses due to
the Rashba spin-orbit (SO) interaction arising from the lack of
structural inversion symmetry in the semiconductor heterostructures. 
The spin of the
electrons can be made to transport coherently and the spin-polarized drain 
whose spin polarization is parallel to the same as source can detect the
transportation of spin. The spin current can be manipulated by controlling
gate voltage since the Rashba SO coupling depends on it.
However, the momentum relaxation due to elastic scattering 
of the electrons from impurity potential leads to spin relaxation and thereby
destroys spin coherence. 
Although SFET has not been achieved yet, the modulation of SO
coupling by gate voltage has been observed in InGaAs/InAlAs and
GaAs/AlGaAs heterostructures\cite{Expt_1,Expt_2,Expt_3}.
This motivates to study the system of electrons with SO interactions.

A remarkable consequence of the SO interaction is the spin-Hall effect
where an electric field induces a transverse out-of-plane spin-current
and thereby a spin imbalance takes place between two opposite
edges of the sample.
Ever since the prediction of intrinsic dissipationless spin-Hall effect  
in hole-doped semiconductor systems\cite{Zhang}, this effect has drawn intense
theoretical as well as experimental activities.
Although Sinova {\it et al} \cite{Pure} 
have predicted universal (independent of the
strength of Rashba spin-orbit interaction) {\it dc}
spin-Hall conductivity (SHC) $\sigma^z_{xy} =e/8\pi$ in a pure 
2DES, the issue of SHC in the presence of non-magnetic
impurities remains highly controversial. A number of analytical works
\cite{SHall,Halperin,Loss} based on Kubo formula and quantum kinetic
equations
suggest that $\sigma^z_{xy}$ vanishes for any amount of elastic disorder in
the diffusive transport regime. 
This dramatic vanishing of SHC in the presence of disorder is argued to
be a more generic phenomenon \cite{Rashba2} as $\sigma^z_{xy}$ also
vanishes for any amount of magnetic field in a pure system.
On the contrary, experiments \cite{Kato,Wunderlich,Sih}
seem to suggest that there is nonequilibrium
spin accumulation in the edges, transverse to the direction of application
of charge current. 
Numerical studies \cite{Ting}
based on the Landauer-Buttiker approach in mesoscopic systems 
with leads cannot shed any light about SHC in the thermodynamic limit because
of the possibility of edge effects near the contacts.
However, a numerical calculation based on the Kubo formula in the
lattice model \cite{Haldane} suggests that SHC is finite in the presence of
disorder, while it vanishes in the thermodynamic limit.
Nomura et al \cite{Nomura1}
have numerically shown that the vanishing of intrinsic
$\sigma^z_{xy}$ is peculiar to the linear-momentum Rashba model which
has been considered in the previous studies. They have shown that for Rashba
model with cubic momentum which is the case for two dimensional hole gas,
the intrinsic $\sigma^z_{xy}$ is nonzero and consequently intrinsic edge-spin
accumulation takes place in the systems of two dimensional hole gases 
\cite{Wunderlich,Nomura2}.
On the other hand, the edge-spin accumulation observed \cite{Kato,Sih}
in the system of two
dimensional electron gas where Rashba model with linear momentum is important,
is argued to be extrinsic by the `skew-scattering' mechanism
\cite{Nagaosa,Halperin2,SDSarma}.
While the issue of intrinsic versus extrinsic spin-Hall effect is not
clear yet,
we show below that the linear Rashba model provides a different 
but robust kind of spin-Hall
effect in the two dimensional electron systems.

In this paper, we calculate the spin-Hall conductivities
in the charge-spin space using Kubo formula in section III.
A charge current
induces an out-of-plane component of spin current in the transverse
direction. 
This usual ac spin-Hall conductivity $\sigma^z_{xy}$ is found to be
the same as in a kinetic equation approach \cite{Halperin}. It however
vanishes in the dc limit \cite{SHall,Halperin,Loss}.
We call this phenomenon as `charge-spin' Hall effect.
Apart from this, we find that
an in-plane spin-polarized current can induce a transverse spin-polarized current 
along transverse spatial direction. 
This phenomenon which we call `spin-spin' Hall effect is a spin-analog
of the conventional charge Hall effect, but in the absence of magnetic
field.
The relevant Hall conductivity of this phenomenon $\sigma^\perp_{xy}$ does not
vanish in the dc limit. We attribute this phenomenon to the diffusion of
spin $s_y$ along $y$-direction due to a source causing conventional
$s_x$ spin-diffusion
along $x$-direction. The corresponding diffusive equation is derived
in section IV. The detailed derivation of diffusive propagator is
available in the Appendix A. 
In Section V, the spin-spin Hall conductivity has been argued to be
robust. Relevance of other type of spin-spin Hall conductivities
have also been discussed. The consequence of Dresselhaus 
SO coupling along with the Rashba coupling on spin-spin Hall conductivities has
been discussed in Appendix B.

\section{Hamiltonian and Green's Functions}

A system of two dimensional non-interacting electrons with Rashba
spin-orbit interaction in the presence of non-magnetic impurities
from which the electrons scatter elastically can be described by
the Hamiltonian
\be
H=H_0+V(\r);\, H_0 = \frac{\p^2}{2m} + \lambda \hat{{\bm \eta}} \cdot \p \, ,
\label{Hamiltonian}
\ee 
where $\p = -i\nabla$ is the momentum operator, $m$ is the effective band mass
of electrons, $\lambda$ is the parameter for the strength of Rashba
spin-orbit interaction, $\hat{\bm \eta} = {\bf n} \times \hat{\bm{\sigma}}$ is
a spin operator with $\hat{\bm{\sigma}}$
being the Pauli matrices, ${\bf n}$ is the unit vector perpendicular 
to the plane of the system, and $V(\r)$ is the disorder potential.
The eigenvalues of $H_0$ are $\epsilon_\k^s = k^2/2m + s\lambda k$
where $k$ is the magnitude of the wavevector $\k$ of the electrons,
$s= \pm$  is the index for spin-split subband $s$. 
The corresponding eigenstates are given by
\be
\psi_\k^s(\r) = e^{i\k \cdot \r} \frac{1}{\sqrt{2}} 
   \lp \begin{array}{c}  e^{i\chi_{\k}/2}  \\
     s e^{-i\chi_{\k}/2}
    \end{array}\rp \, ,
\ee
where $\chi_\k = \pi/2 -\theta$ with $\theta$ being the angle of $\k$
with $x$-axis. 
The Fermi momenta and the density of states at Fermi
energy, $\epsilon_F$, in the two subbands are $k_F^s
=\sqrt{k_F^2+m^2\lambda^2}-sm\lambda$ and $\nu^s = \nu (1-s\frac{m\lambda}{
\sqrt{k_F^2+m^2\lambda^2}})$
respectively, where $\nu = m/2\pi$ is the density
of states for each spin direction and $k_F = \sqrt{2m\epsilon_F}$ is the
Fermi momentum in the absence of SO interaction.
The charge and spin currents are $\hat{\j}_\alpha = e_\alpha \frac{1}{2}
\lf\hat{\sigma}_\alpha ,\, \hat{{\bf v}}_\k \rf$ where $\hat{{\bf v}}_\k
= \frac{\k}{m} \hat{\sigma}_0 +\lambda \hat{ {\bm \eta}}$ is the velocity and
$\hat{\sigma}_0$ is the unit matrix .
Here $\alpha = 0$--3 represents charge and three spin directions respectively
and $e_0 = e$ (charge of the electrons), $e_1=e_2=e_3 = 1/2$ (spin of the
electrons).

We next assume random disorder potential with the
configurational average  such that $\langle V(\r)\rangle =0$
and $\langle V(\r) V(\r')\rangle = \gamma \delta (\r -\r')$.
The elastic life time of the electrons will then be
$\tau = 1/(2\pi\gamma\nu)$. Therefore, the retarded (advanced)
Green's function for the electrons with energy $\epsilon$ and
wavevector $\k$ can be expressed as 
\be
\hat{\G}^{R,A}(\k ,\, \epsilon) = \frac{1}{2}\sum_{s=\pm}
  \frac{\hat{\sigma}_0 + s\hat{\eta}_\k}{\epsilon - \xi^s_\k \pm i/2\tau}
  \equiv \sum_{s=\pm} \hat{\G}^{R,A}_s(\k,\, \epsilon) \, ,
\label{Green}
\ee
where $\hat{\eta}_\k = \hat{\bm \eta}\cdot \k /k$ is the projection of
the spin operator into the direction of $\k$ and
$\xi_\k^s = \epsilon_\k^s-\epsilon_F$.

\section{Hall Conductivities}

The spin-Hall conductivities $\sigma^z_{xy}$, $\sigma^\perp_{xy}$, and
$\sigma^\parallel_{xy}$ correspond to transverse 
induction of out-of-plane (z-axis) spin-current due to a charge current,
transverse (in plane) spin-current due to a spin-polarized current, and
parallel spin-current due to a spin-polarized current respectively.
These conductivities
can be obtained using Kubo formula:
\bea
\sigma^z_{xy}(\omega) &=& \frac{1}{4\pi}\lp \Pi^{03}_{xy}+\Pi^{30}_{xy} \rp \, ,
\label{Conductivity} \\
\sigma^\perp_{xy}(\omega) &=& 
\frac{1}{4\pi}\lp \Pi^{12}_{xy}+\Pi^{21}_{xy} \rp \, , 
\label{spinperp}\\
\sigma^\parallel_{xy}(\omega) &=& \frac{1}{4\pi}\lp \Pi^{11}_{xy}+\Pi^{22}_{xy} \rp \, ,
\label{spinparal}
\eea
where retarded off-diagonal current-current correlation functions 
can be written as
\be
 \Pi^{\alpha\beta}_{xy} (\omega) 
    =   \int \frac{d\k}{(2\pi)^2}\, \text{Tr} 
  \lp    \hat{j}^x_\alpha(\k) 
  \hat{\G}^A(\k , 0) \, 
       \lb \hat{j}^y_\beta (\k) +\hat{J}^y_\beta (\omega) \rb 
       \hat{\G}^R(\k ,  \omega)\rp 
\label{Pi}
\ee
Here the current $\hat{\j}_\beta(\k)$ corresponds to the bare vertex. 
This vertex gets corrected through
$\hat{\J}_\beta$ which is the self-consistent solution of the equation
\be
\hat{\J}_\beta (\omega) = \gamma\int\frac{d\k'}{(2\pi)^2}\hat{\G}^A(\k',0)
              \lb \hat{\j}_\beta(\k') + \hat{\J}_\beta(\omega) \rb 
                \hat{\G}^R(\k',\omega) \, .
\label{Ladder}
\ee 
It describes summation over infinite number of
the ladder diagrams in a diagrammatic approach.

To solve Eq.~(\ref{Ladder}) and evaluate Eq.~(\ref{Pi}), we need to
consider integrations, {\it e.g.},
$$\int \frac{d\k}{(2\pi)^2} \hat{\G}^A_s(\k,0)\hat{\G}^R_{s'}(\k,\omega) $$
which we calculate using the approximation
\be
\int \frac{d\k}{(2\pi)^2} \Rightarrow
       \int_0^{2\pi}\frac{d\theta}{2\pi} 
\lb \nu^s \int^\infty_{-\infty}
        d\xi_\k^s \delta_{ss'} 
  +\nu\int^\infty_{-\infty} d\xi_\k (1-\delta_{ss'})\rb
\label{Integration}
\ee  
{\it i.e.}, the integrations are performed around momentum $k_F^s$ for
the intraband contribution while the interband contributions are obtained
through the expansion around $k_F$ and the variable of integration
for the latter
is $\xi_\k = k^2/2m-\epsilon_F$.  
This approximation is, however, neither biased towards any of the subbands
nor negligent of the existence of those. This attention is important
towards our result.
The self-consistent solutions are thus straightforward and we find as
\bea
\hat{\J}_0(\omega) &=& - \lambda\,\frac{2\delta^2}
  {2\delta^2(1-2i \omega \tau) -i\omega\tau(1-i\omega\tau)^2} \hat{{\bm \eta}} 
\label{j0}\\
\hat{\J}_1(\omega) &=& -\lp \frac{k_F}{2m}\rp\, \frac{\delta}{4\delta^2
                        -i\omega\tau (1-i\omega\tau)} 
                          \hat{\sigma}_3\hat{{\bm x}}
\label{j1}\\
\hat{\J}_2(\omega) &=& -\lp \frac{k_F}{2m}\rp\, \frac{\delta}{4\delta^2
                        -i\omega\tau (1-i\omega\tau)} 
                    \hat{\sigma}_3\hat{{\bm y}}
\label{j2}\\
\hat{\J}_3(\omega) &=& \lp \frac{k_F}{2m}\rp\, \frac{\delta (1-i\omega\tau)}
  {2\delta^2(1-2i \omega \tau) -i\omega\tau(1-i\omega\tau)^2} \hat{{\bm \sigma}} 
\label{j3}
\eea
for an arbitrary value of the parameter $\delta = \lambda k_F\tau$. 
While $\hat{\J}_0$
and $\hat{\J}_3$ are due to both the intraband and the interband contributions,
$\hat{\J}_1$ and $\hat{\J}_2$ are due to interband contribution only. 
This means $\hat{\J}_1$ and $\hat{\J}_2$ exist only because of the formation
of two subbands. In fact all of these vanish for
$\lambda = 0$, in consistence with the fact that the charge transport time and
the momentum relaxation time are the same (no vertex correction)
in a two dimensional system with short range scatterers.

Using equations (\ref{Conductivity})--(\ref{j3})
 we find the Hall conductivities:
\bea
\sigma^z_{xy}(\omega) &=& \lp \frac{e\delta^2}{2\pi}\rp \frac{i\omega\tau}
    {2\delta^2(1-2i \omega \tau) -i\omega\tau(1-i\omega\tau)^2} \, ,
         \label{Sigmacs}\\
\sigma^\perp_{xy}(\omega) &=& -\lp \frac{\Delta}{16\pi}\rp \frac{4\delta^2}{[4\delta^2
       -i\omega\tau(1-i\omega\tau)][1-i\omega\tau]}  \, , 
             \label{Sigmass} \\
\sigma^\parallel_{xy} (\omega) &=& 0 \, .
\label{Sigmap}
\eea
The spin-Hall conductivity $\sigma^z_{xy}(\omega)$ in Eq.~(\ref{Sigmacs})
is identical to those derived from 
the quantum kinetic equation by Mishchenko {\it et al}\cite{Halperin}
and in an approach of Kubo formula by Chalaev and Loss \cite{Loss}. 
At $\omega =0$, the
vertex correction term dramatically cancels \cite{SHall,Halperin,Loss}
the bare contribution in $\sigma^z_{xy}$.
The vanishing of $\sigma^z_{xy}$ even in the weak localization regime
\cite{Loss} warns the presence of an exact property of the system under
consideration. From a general argument, Chalaev and Loss \cite{Loss}
have shown $\sigma^z_{xy}$ vanishes exactly for any finite amount of disorder.
Since the spin is not conserved, the spin-Hall conductivity may not be
directly related \cite{Rashba2} to the observed \cite{Kato,Wunderlich,Sih}
spin accumulation at the transverse edges.

The most important result in this paper is the finding of the conductivity
$\sigma^\perp_{xy}(\omega)$. This suggests that an in-plane spin polarized current
induces a transverse spin current with in-plane spin polarization but
perpendicular to the former. 
If the source of the spin current is polarized along $x$-axis, then a
detector, placed transverse to the direction of the applied current,
with polarization along $\pm y$-axis can detect this novel current.
Magnetic semiconductors may be used for spin injection as well as detection
of the spin-polarized currents 
as described by Fiederling et al \cite{Molenkamp} in a light-emitting diode.
Further techniques \cite{Kato,Wunderlich,Sih}
developed for the detection of spin-accumulation in spin-Hall
effect may also be useful for observing spin-spin-Hall effect.
Unlike the conventional spin Hall conductivity $\sigma^z_{xy}$, this new
spin-spin Hall conductivity is finite at $\omega = 0$ and is given by
\be
\sigma^\perp_{xy}(\omega = 0) = -\frac{\Delta}{16\pi} \, ,
\label{Sigmadc}
\ee
independent of the Rashba coupling parameter $\lambda$.
However from the equation (\ref{Sigmass}) we see that
 $\sigma^\perp_{xy}(\omega = 0) \neq 0$ only when $\lambda \neq 0$.
In other words, $\lambda \rightarrow 0$ before $\omega \rightarrow 0$
is logically the correct limiting procedure for dc $\sigma^\perp_{xy}$ in the
absence of spin-orbit interaction.
The bare and vertex correction contribution to $\sigma^\perp_{xy}$ in 
Eq.~(\ref{Sigmadc}) are $-\frac{\Delta}{16\pi}\lp
\frac{4\delta^2}{1+4\delta^2}\rp$ and $-\frac{\Delta}{16\pi}\lp
\frac{1}{1+4\delta^2}\rp$ respectively. Clearly the dominant contribution
is due to the vertex correction. The vanishing $\sigma^\parallel_{xy}$ 
in Eq.~(\ref{Sigmap}) implies
zero parallel spin-polarized current in the transverse direction.

\section{Spin Diffusion} 

We now derive the spin-diffusive equation for understanding the above
novel spin-spin Hall effect.
The retarded charge and spin density correlation functions at frequency
$\omega$ and momentum $\q$  can be written as
\be
\chi_{\alpha\beta}(\q ,\omega) = \frac{i\omega}{2\pi}\int\frac{d\k}{(2\pi)^2}
 \text{Tr} \lb \hat{\sigma}_\alpha \hat{\G}^A(\k , 0)\hat{M}_\beta
 \hat{\G}^R_\beta(\k+\q, \omega) \hat{\sigma}_\beta\rb
\label{Correlation}
\ee
where $\alpha ,\, \beta =$ 0--3 corresponding to charge and three spin
directions, $\hat{\G}^R_\beta (\k , \epsilon) = \hat{\sigma}_\beta
\hat{\G}^R(\k , \epsilon)\hat{\sigma}_\beta$, and $\hat{M}_\beta
= \sum_{j=0}^\infty \hat{m}^{(j)}_\beta$ with $\hat{m}^{(0)}_\beta
=\hat{\sigma}_0$ and 
\be
\hat{m}^{(j)}_\beta = \gamma \int \frac{d\k'}{(2\pi)^2}\hat{\G}^A(\k' ,0)
\hat{m}^{(j-1)}_\beta \hat{\G}_\beta^R(\k'+\q , \omega) \, ; j \geq 1
\label{mbeta}
\ee
It is very easy to check that the Eq.~(\ref{Correlation}) includes 
infinite number of ladder diagrams
for the vertex correction. Although the expression of $\chi_{\alpha\beta}$
(\ref{Correlation}) appears to be uncommon, we find it to be very convenient
way to express for all orders of diagrams since we need to maintain the order
of matrices inside the trace. 
The expression of $\chi_{\alpha\beta}$ in (\ref{Correlation}) is formal.
We however evaluate it for small $q$ and $\omega$ such that
$\omega \tau \ll 1$ and $ q \ll k_F$, and using the approximation in
the expression (\ref{Integration}) for the integration over electron momenta. 
We iteratively calculate the coefficients of $\hat{\sigma}$'s in $\hat{m}_\beta$'s
(\ref{mbeta}) and take sum of geometrical series of the coefficients.
The readers interested in the details of the calculation may look into the
Appendix A.

Expressing 
$\hat{m}_\beta^{(1)}
 = \sum_{\alpha = 0}^3 m_{\beta\alpha}\hat{\sigma}_\alpha$ and
\be
\chi_{\alpha\beta} = 2i\nu\omega\tau  \left( {\cal D}_{\alpha\beta}
     - \delta_{\alpha\beta} \right) \,  
\ee
in a standard form,
we find the inverse diffusion propagator to be
\be
{\cal D}^{-1} = \lb \begin{array}{cccc}
 1-m_{00} & -m_{11} & -m_{22} & 0  \\
 -m_{01} & 1-m_{10} & \frac{m_{21}m_{32}}{1-m_{30}} & -i\,m_{32} \\
 -m_{02} & \frac{m_{12}m_{31}}{1-m_{30}} & 1-m_{20} &  i\,m_{31} \\
0 & i\,m_{12} & -i\,m_{21} & 1-m_{30} \end{array} \rb
\label{Propagator}
\ee
where 
$m_{00} = 1 +\tau (i\omega - D\q^2), \, m_{10}=m_{20}=m_{00}-\tau /
\tau_s^\parallel,
\, m_{30}=m_{00}-\tau / \tau_s^\perp$ with diffusion constant
$D = \Delta/m$, and $\tau_s^\perp = \tau(1+4\delta^2)/4\delta^2$ and 
$\tau_s^\parallel =
\tau(1+4\delta^2)/2\delta^2$ being the out-of-plane and in-plane
spin relaxation times respectively, $m_{01}=m_{11}=
2iq_y\delta^3/k_F(1+4\delta^2)$, $m_{02} = m_{22} = 
 - (q_x/q_y)m_{01}$, $m_{12} = m_{32}= -4q_x\delta\Delta/k_F(1+4\delta^2)^2$
 and $m_{21} = m_{31} = - (q_y/q_x) m_{12}$. 
We thus find ${\cal D}^{-1}_{21} = -2q_xq_y\Delta\tau/m(1+4\delta^2)^3$ for
$\omega = 0$. 
Clearly ${\cal D}^{-1}_{21}$ and ${\cal D}^{-1}_{12}$ are nonzero 
for small $\delta$, contrary 
to the recent result of Burkov et al\cite{Burkov}.
(See Appendix A for details). This is because the 
spin-relaxation provides an energy cut-off.
These are indeed the
responsible terms for driving transverse spin in the transverse
direction as we shall show below. However, the other
components of ${\cal D}^{-1}$ are in agreement with Ref.~\onlinecite{Burkov}.

Transformation of the inverse diffusive propagator (\ref{Propagator}) into the 
real space and time leads to the transport equations for spins with planer
projection (neglecting gradients of charge and out-of-plane 
component of spin densities):
\bea
\frac{\partial S_x}{\partial t} = \left( D\bm{\nabla}^2 
    -\frac{1}{\tau_s^\parallel}
    \right) S_x - \frac{2D}{(1+4\delta^2)^3}
             \partial_x\partial_yS_y  +I_x & & 
\label{Trans1}\\
\frac{\partial S_y}{\partial t} = \left( D\bm{\nabla}^2 
    -\frac{1}{\tau_s^\parallel}
    \right) S_y - \frac{2D}{(1+4\delta^2)^3}
       \partial_x\partial_yS_x  +I_y &&
\label{Trans2}
\eea
where $S_x (S_y)$ is spin density along $x (y)$ direction and $x (y)$-component
of the spin current $I_x (I_y)$ is injected into the system. 
Although the coefficients of $\partial_x\partial_yS_y$ in Eq.~(\ref{Trans1})
and $\partial_y\partial_xS_x$ in Eq.~(\ref{Trans2}) appear to be nonzero for
$\delta =0$, these coefficients in fact vanish strictly at $\delta =0$. The
spin diffusive equations (\ref{Trans1}) and (\ref{Trans2}) are valid for
$\omega \tau \ll 4\delta^2$. 
Clearly, the injected spin
current $I_x (I_y)$ not only transports
the $S_x (S_y)$ 
component of spin along the direction of the current but also drives
the $S_y (S_x)$ component of spin along the transverse direction.  
The finite dc spin-spin Hall conductivity (\ref{Sigmadc}) including sign
arises as a manifestation of this fact.

\section{Discussion and Summary}

Since, due to the precession, the spin is not conserved \cite{Rashba2,Burkov},
the definition of
spin current is not unique. We, however, have considered generally accepted
definition of the spin-current operator, i.e., the symmetrization of the
product of spin operator and the group velocity of electrons. 
Apart from this spin-current
which depends on the translational motion of spin, there is a current associated
with the rotational motion\cite{Xie}. Albeit the definition of current
is not unique, the arbitrariness lies only in the terms proportional to the
SO coupling parameter $\lambda$. The kinetic parts\cite{Burkov}
(which are proportional to the momentum of electrons and independent
of $\lambda$) of the spin-currents are, however, unique. 
The substantial contribution in $\sigma^\perp_{xy}$ 
which arises due to vertex correction (\ref{Sigmadc}) is entirely due to the
kinetic part of the spin currents, and it is not proportional to any
power of $\lambda$.
We therefore believe that
the `spin-spin' Hall effect is robust in the system of two dimensional electron
gas with the Rashba spin-orbit interaction.

We have calculated $\sigma_{xy}^\perp$ and $\sigma_{xy}^\parallel$
which suggest spin-spin Hall conductivities when the induced spin-current
will have respective 
spin-polarization perpendicular and parallel to the spin-polarization
of applied spin-current. 
These spin-polarizations are in the plane of the 2DES. 
While the mechanism for usual spin-Hall effect
in this system is mainly the momentum dependent spin-precession
along with the spin relaxation,
the mechanism behind spin-spin Hall effect is spin-diffusion.
If the spin-polarization of the applied spin-current is out of the plane
of the system, we find the corresponding components of off-diagonal
current-current correlation functions (\ref{Pi}) vanish:
$\Pi_{xy}^{13}+\Pi_{xy}^{31}
= \Pi_{xy}^{23} + \Pi_{xy}^{32} = \Pi_{xy}^{11}+\Pi_{xy}^{33} =
\Pi_{xy}^{22}+\Pi_{xy}^{33} = 0$. 
In other words, the spin-spin Hall effect is possible {\em only} when
the applied spin-current's polarization is in the plane of
the system.

In summary, we predict that an in-plane spin-polarized current can induce
a transversed spin-polarized current along transverse direction. This phenomenon
which we call `spin-spin' Hall effect is a spin analog of conventional Hall
effect, but with no magnetic field. The reason behind this phenomenon is 
the transverse spin diffusion along transverse direction due to the application
of a source causing conventional spin diffusion.

\section*{Acknowledgment}

 One of us (SSM) thanks Jainendra K. Jain, T. V. Ramakrishnan
 and Sankar D. Sarma for discussions.

\appendix

\widetext
\section{Diffusive Propagator}

In this appendix, we present detailed calculation of the charge-spin
density correlation function, and hence inverse diffusive propagator and compare
with previous result \cite{Burkov}. The retarded
charge-spin density correlation function can be represented by a sum of infinite
number of ladder diagrams:
\be
\chi_{\alpha\beta}(\q ,\omega) = \sum_{j = 0}^\infty \chi_{\alpha\beta}^{(j)}
(\q ,\omega),
\ee
where (j) represents the order of the diagrams.
Explicitly these are given by
\bea
\chi_{\alpha\beta}^{(0)}(\q ,\omega) &=&
\frac{i\omega}{2\pi}\text{Tr}\lb 
\int\frac{d\k}{(2\pi)^2} \hat{\sigma}_\alpha \hat{\G}^A(\k , 0)
\hat{\sigma}_\beta  \hat{\G}^R(\k+\q, \omega)\rb \\
\chi_{\alpha\beta}^{(1)}(\q ,\omega) &=& \frac{i\omega}{2\pi}
\text{Tr} \lb 
\gamma  \int \frac{d\k}{(2\pi)^2}
\int\frac{d\k_1}{(2\pi)^2} \hat{\sigma}_\alpha \hat{\G}^A(\k , 0) \right.
 \nonumber\\
&\times & \left. \hat{\G}^A(\k_1 ,0)
\hat{\sigma}_\beta \hat{\G}^R(\k_1+\q , \omega)\hat{\G}^R(\k+\q, \omega)
\rb 
\eea
and so on. In the expression of $\chi_{\alpha\beta}$, the contribution at
$\omega =0$ and $q=0$ for diagonal terms has been ignored.
It is crucial to maintain the relative positions  of the matrices
inside the trace while performing the infinite series. To express all the
orders in a compact form and to calculate those,
we find it convenient to shift  $\hat{\sigma}_\beta$
to the extreme right; but in doing so,  $\hat{\G}^R (\k+\q, \omega)$  will
be modified to  $\hat{\G}^R_\beta
(\k+\q,\omega)=\hat{\sigma}_\beta\hat{\G}^R (\k+\q, \omega)
\hat{\sigma}_\beta$. We thus find
\bea
\chi_{\alpha\beta}^{(0)}(\q ,\omega) &=& \frac{i\omega}{2\pi}\text{Tr}\lb 
\hat{\sigma}_\alpha \int\frac{d\k}{(2\pi)^2}\hat{\G}^A(\k , 0)
\hat{\G}^R_\beta(\k+\q, \omega)\hat{\sigma}_\beta\rb 
\label{chi0} \\
\chi_{\alpha\beta}^{(1)}(\q ,\omega) &=& \frac{i\omega}{2\pi}
\text{Tr}\lb 
\hat{\sigma}_\alpha\int\frac{d\k}{(2\pi)^2} \hat{\G}^A(\k , 0)
\lf\gamma \int \frac{d\k_1}{(2\pi)^2} \right. \right. \nonumber \\
& & \left. \left. \hat{\G}^A(\k_1 ,0)
\hat{\G}^R_\beta(\k_1+\q , \omega)\rf 
\hat{\G}^R_\beta(\k+\q, \omega)
\hat{\sigma}_\beta\rb \\
\chi_{\alpha\beta}^{(2)}(\q ,\omega)&=&\frac{i\omega}{2\pi}
\text{Tr} \lb
\hat{\sigma}_\alpha\int\frac{d\k}{(2\pi)^2} \hat{\G}^A(\k , 0)
\lp \gamma \int \frac{d\k_2}{(2\pi)^2}\hat{\G}^A(\k_2 ,0) \right. \right.
\nonumber\\
&&\left.\left. \lf\gamma \int \frac{d\k_1}{(2\pi)^2}\hat{\G}^A(\k_1 ,0)
\hat{\G}^R_\beta(\k_1+\q , \omega)\rf 
\hat{\G}^R_\beta(\k_2+\q, \omega) \rp
\hat{\G}^R_\beta(\k+\q, \omega)
\hat{\sigma}_\beta \rb
\label{chi2}
\eea
and so on.
Defining
\be
\hat{m}^{(j)}_\beta (\q,\omega)= \gamma \int \frac{d\k'}{(2\pi)^2}
\hat{\G}^A(\k' ,0)
\hat{m}^{(j-1)}_\beta (\q , \omega)\hat{\G}_\beta^R(\k+\q , \omega)
\,\, ;\, j \geq 1
\label{mdef}
\ee
with $\hat{m}^{(0)}_\beta=\hat{\sigma}_0$, and summing over all orders 
(the lowest three orders given in Eqs.~(\ref{chi0})--(\ref{chi2})),
we rewrite the retarded charge-spin density correlation function as
\be
\chi_{\alpha\beta}(\q ,\omega) =
 \frac{i\omega}{2\pi}\text{Tr} \lb 
\hat{\sigma}_\alpha \int\frac{d\k}{(2\pi)^2}\hat{\G}^A(\k , 0)
\lp\sum_{j = 0}^\infty\hat{m}^{(j)}_\beta(\q,\omega)\rp
\hat{\G}^R_\beta(\k+\q, \omega)\hat{\sigma}_\beta\rb
\label{xif}
\ee
Expressing 
\bea
\hat{m}^{(1)}_\beta(\q,\omega)&=&\gamma\int \frac{d\k_1}{(2\pi)^2}
\hat{\G}^A(\k_1 ,0)\hat{\G}^R_\beta(\k_1+\q , \omega)\nonumber \\
&\equiv&m_{\beta0}\hat{\sigma}_0+m_{\beta1}\hat{\sigma}_1+m_{\beta2}\hat{\sigma}_2
+m_{\beta3}\hat{\sigma}_3
\eea
with $\beta$ varying from 0 to 3, we obtain 16 components of $m_{\beta\alpha}$
which are
$m_{00}=1 +\tau (i\omega - D\q^2)$, 
$m_{10}=m_{20}=m_{00}-\frac{2\delta^2}{1+4\delta^2}$, 
$m_{30}=m_{00}-\frac{4\delta^2}{1+4\delta^2}$,
$m_{01}=m_{11}= 2i\frac{q_y}{k_F}\frac{\delta^3}{(1+4\delta^2)}$,
$m_{02} = m_{22} = - \frac{q_x}{q_y}m_{01}$,
$m_{12} = m_{32}= -\frac{q_x}{k_F}\frac{4\delta\Delta}{(1+4\delta^2)^2}$,
$m_{21} = m_{31} = - \frac{q_y}{q_x}m_{12}$,
$m_{03}= m_{33}= 0$ , $m_{13}$ and $m_{23}$
are ${\cal O} \lp \frac{q^2}{k_F^2},\delta^2 \rp$ and are very small compared 
to the others.
Here Diffusion constant $D=\frac{1}{2}v_F^2\tau$,
dimensionless Rashba coupling strength $\delta=\lambda k_F\tau$
and $\Delta=\epsilon_F\tau$.

One has to now sum four infinite series of matrices, one for each value of
$\beta$, to find out the matrix 
$\sum_{j = 0}^\infty\hat{m}_\beta^{(j)} $, using
~(\ref{mdef}) iteratively. 
We show the explicit calculation of $\sum_{j = 0}^\infty\hat{m}_1^{(j)}$ below,
others may be obtained following the same procedure.

From Eq.~(\ref{mdef}), we find
\bea
\hat{m}_1^{(2)}&=&\gamma \int \frac{d\k'}{(2\pi)^2}
\hat{\G}^A(\k' ,0)
\hat{m}^{(1)}_1(\q,\omega) \hat{\G}_1^R(\k'+\q , \omega)\nonumber\\
&=&\gamma \int \frac{d\k'}{(2\pi)^2}\lb 
m_{10}\hat{\G}^A\hat{\sigma}_0
\hat{\G}^R_1+m_{11}\hat{\G}^A\hat{\sigma}_1\hat{\G}^R_1
+m_{12}\hat{\G}^A
\hat{\sigma}_2\hat{\G}^R_1+m_{13}\hat{\G}^A\hat{\sigma}_3\hat{\G}^R_1 \rb
\nonumber\\
&=&\gamma \int \frac{d\k'}{(2\pi)^2}\lb 
m_{10}\hat{\G}^A
\hat{\G}^R_1\hat{\sigma}_0+m_{11}\hat{\G}^A\hat{\G}^R_0\hat{\sigma}_1
+m_{12}\hat{\G}^A
\hat{\G}^R_3\hat{\sigma}_2+m_{13}\hat{\G}^A\hat{\G}^R_2\hat{\sigma}_3\rb 
\nonumber \\
&=& m_{10}\hat{m}_1^{(1)}\hat{\sigma}_0+m_{11}\hat{m}_0^{(1)}\hat{\sigma}_1+
m_{12}\hat{m}_3^{(1)}\hat{\sigma}_2+m_{13}\hat{m}_2^{(1)}\hat{\sigma}_3
\label{m2}
\eea
Putting $\hat{m}_\beta^{(1)}$ in (\ref{m2}) 
and collecting the coefficients of each Pauli matrix, we obtain
\bea
\hat{m}_1^{(2)}&=&\hat{\sigma}_0\lb ( m_{10})^2+ m_{11} m_{01}
+ m_{12} m_{32}+ m_{13} m_{23}\rb 
+\hat{\sigma}_1\lb m_{11} m_{10}
+ m_{11} m_{00}+ i m_{13} m_{22}\rb  \nonumber \\
&&+\hat{\sigma}_2\lb m_{12} m_{10}
+ m_{12} m_{30}- i m_{13} m_{21}\rb 
+\hat{\sigma}_3\lb m_{13} m_{10}
-i m_{11} m_{02}+i m_{12} m_{31}+ m_{13} m_{20}\rb \nonumber\\
&\approx&\hat{\sigma}_0\lb ( m_{10})^2\rb 
+\hat{\sigma}_1\lb  m_{11}( m_{10}+m_{00})\rb 
 +\hat{\sigma}_2 \lb m_{12}( m_{10}+m_{30})\rb
+\hat{\sigma}_3 \lb i m_{12}m_{31}\rb  \, ,
\eea
where the smaller terms are neglected in the last expression. Note that, each term
in coefficient of $\hat{\sigma}_3$ is ${\cal O}(q^2/k_F^2)$, but the one which is
kept has much larger coefficient than the others. Similarly,  
\bea
\hat{m}_1^{(3)} &\approx& \hat{\sigma}_0( m_{10})^3+\hat{\sigma}_1 m_{11}
\lb (m_{10})^2+(m_{00})^2+m_{10}m_{00}\rb  \nonumber \\
& & +\hat{\sigma}_2\lb (m_{10})^2+(m_{30})^2+m_{10}m_{30}\rb
+\hat{\sigma}_3 \lb i m_{12}m_{31}\lp m_{10}+m_{20}+m_{30} \rp\rb
\eea
One can anticipate the higher orders now. Summing all the orders,
\bea
\sum_{j = 0}^\infty\hat{m}_1^{(j)} &=& \hat{\sigma}_0 \frac{1}{1-m_{10}}
+\hat{\sigma}_1\frac{m_{11}}{(1-m_{10})(1-m_{00})} \nonumber \\
& &+\hat{\sigma}_2\frac{m_{12}}{(1-m_{10})(1-m_{30})}
+\hat{\sigma}_3\frac{i m_{12}m_{31}}{(1-m_{10})(1-m_{20})(1-m_{30})}
\label{sum1}
\eea
We similarly find
\bea
\sum_{j = 0}^\infty\hat{m}_0^{(j)} &=& \hat{\sigma}_0 \frac{1}{1-m_{00}}
+\hat{\sigma}_1\frac{m_{01}}{(1-m_{00})(1-m_{10})} 
+\hat{\sigma}_2\frac{m_{02}}{(1-m_{00})(1-m_{20})} \\
\sum_{j = 0}^\infty\hat{m}_3^{(j)} &=& \hat{\sigma}_0 \frac{1}{1-m_{30}}
+\hat{\sigma}_1\frac{m_{31}}{(1-m_{20})(1-m_{30})} 
+\hat{\sigma}_2\frac{m_{32}}{(1-m_{10})(1-m_{30})} \\
\sum_{j = 0}^\infty\hat{m}_2^{(j)} &=& \hat{\sigma}_0 \frac{1}{1-m_{20}}
+\hat{\sigma}_1\frac{m_{21}}{(1-m_{20})(1-m_{30})} \nonumber \\
& &+\hat{\sigma}_2\frac{m_{22}}{(1-m_{20})(1-m_{00})}
+\hat{\sigma}_3\frac{-i m_{21}m_{32}}{(1-m_{10})(1-m_{20})(1-m_{30})}
\label{sum2}
\eea

It is now straightforward to calculate $\chi_{\alpha \beta}$ 
using Eqs.~(\ref{xif}), and (\ref{sum1})--(\ref{sum2}):
\be
\hat{\chi}(\q ,\omega )=2\nu i\omega\tau\lb \begin{array}{cccc}
\frac{m_{00}}{1-m_{00}}& \frac{m_{11}}{(1-m_{00})(1-m_{10})}&
\frac{m_{22}}{(1-m_{00})(1-m_{20})}&0\\
& & & \\
\frac{m_{01}}{(1-m_{00})(1-m_{10})}&\frac{m_{10}}{1-m_{10}}&
-i\frac{m_{21}m_{32}}{(1-m_{10})(1-m_{20})(1-m_{30})}&
i\frac{m_{32}}{(1-m_{10})(1-m_{30})}\\
& & & \\
\frac{m_{02}}{(1-m_{00})(1-m_{20})}&
i\frac{m_{12}m_{31}}{(1-m_{10})(1-m_{20})(1-m_{30})}&\frac{m_{20}}{1-m_{20}}&
-i\frac{m_{31}}{(1-m_{20})(1-m_{30})}\\
& & & \\
0&-i\frac{m_{12}}{(1-m_{10})(1-m_{30})}&i\frac{m_{21}}{(1-m_{20})(1-m_{30})}&
\frac{m_{30}}{1-m_{30}}\end{array}\rb
\label{chimatrix}
\ee 
The diffusive propagator ${\cal D}_{\alpha\beta}$ is connected with charge-spin 
density correlation function by the relation
\be
\chi_{\alpha\beta}=2\nu i\omega\tau(D_{\alpha\beta}-\delta_{\alpha\beta})
\ee
We thus find the inverse diffusive propagator to be
\be
{\cal D}^{-1} = \lb \begin{array}{cccc}
 1-m_{00} & -m_{11} & -m_{22} & 0  \\
 -m_{01} & 1-m_{10} & \frac{m_{21}m_{32}}{1-m_{30}} & -i\,m_{32} \\
 -m_{02} & \frac{m_{12}m_{31}}{1-m_{30}} & 1-m_{20} &  i\,m_{31} \\
0 & i\,m_{12} & -i\,m_{21} & 1-m_{30} \end{array} \rb
\label{diff1}
\ee

Instead of calculating $\hat{\chi}(\q ,\omega)$ explicitly as we have done
above (\ref{chimatrix}), if one defines
\be
\chi_{\alpha\beta}(\q ,\omega) = i\nu\omega\tau {\cal I}_{\alpha \delta}
\overline{{\cal D}}_{\delta\beta}
\label{chibar}
\ee
with
\bea
{\cal I}_{\alpha \beta} &=& \frac{\gamma}{2}\text{Tr} \lb 
\int\frac{d\k'}{(2\pi)^2} \hat{\sigma}_\alpha \hat{\G}^A(\k' , 0)
\hat{\sigma}_\beta  \hat{\G}^R(\k'+\q, \omega)\rb \\
&=& \frac{1}{2}\text{Tr} \lb \sigma_\alpha \hat{m}^{(1)}_\beta \sigma_\beta \rb
\eea
and 
\be
\overline{{\cal D}}^{-1}_{\alpha\beta} = \delta_{\alpha\beta} - {\cal I}_{\alpha
\beta}
\label{dbar}
\ee
as in Ref.~\onlinecite{Burkov}, one finds the inverse diffusive propagator
to be
\be
\overline{{\cal D}}^{-1} = \lb \begin{array}{cccc}
 1-m_{00} & -m_{11} & -m_{22} & 0  \\
 -m_{01} & 1-m_{10} & im_{23} & -i\,m_{32} \\
 -m_{02} & -im_{13} & 1-m_{20} &  i\,m_{31} \\
0 & i\,m_{12} & -i\,m_{21} & 1-m_{30} \end{array} \rb
\label{diff2}
\ee
Clearly ${\cal D}^{-1}_{12} \neq \overline{{\cal D}}^{-1}_{12}$ and 
${\cal D}^{-1}_{21} \neq \overline{{\cal D}}^{-1}_{21}$ but the other
components of ${\cal D}^{-1}$ and $\overline{{\cal D}}^{-1}$ are same. 
While $ \overline{{\cal D}}^{-1}_{12}$ and 
$\overline{{\cal D}}^{-1}_{21}$ are very
small since $m_{13}$ and $m_{23}$ are small due to their quadratic
dependence of the small parameter $\delta$,  ${\cal D}^{-1}_{12}$
 and ${\cal D}^{-1}_{21}$ are not small because these are independent of
$\delta$ in the limit of small $\delta$. These are indeed the important terms
for transverse spin diffusion and hence spin-spin Hall effect.
The way of obtaining inverse diffusive propagator using combined expressions
(\ref{chibar})--(\ref{dbar}) is incorrect
because the noncommutative nature of Pauli matrices is not correctly 
taken care of as we have seen from our explicit calculation of $\hat{\chi}$ in
Eq.~(\ref{chimatrix}).

\section{Role of Dresselhaus Spin-Orbit Coupling}

Due to the bulk inversion asymmetry, the 2DES will also have Dresselhaus
spin-orbit coupling. The strength of this coupling is however much
smaller than the Rashba coupling.
This appendix contains the derivation of the $dc$ spin-spin Hall conductivities 
in presence of both Rashba and Dresselhaus spin-orbit couplings. 
The Hamiltonian in this case is
\be
H=H_0+V(\r)   ;\, H_0 = \frac{\p^2}{2m} + \lambda\lp\hat{\sigma}_1p_y
-\hat{\sigma}_2p_x\rp +\beta\lp\hat{\sigma}_1p_x-\hat{\sigma}_2p_y\rp 
\ee
with the eigenvalues
\be
\epsilon_k^s=\frac{\k^2}{2m}+s\lambda_{RD}k ; \,
    \lambda_{RD}(\theta)=\sqrt{\lambda^2+\beta^2+4\lambda\beta \sin{\theta} 
\cos{\theta}}
\ee
where $\beta$ is the Dresselhaus SO coupling strength.
The energy spectrum now depends on modified SO coupling strength $\lambda_{RD}
(\theta)$. 
Fermi surface of both the branches are anisotropic here and the corresponding
Fermi momenta are $k_F^s = -sm\lambda_{RD} + \sqrt{k_F^2+m^2\lambda_{RD}^2}$.

The retarded (advanced) Green's functions for the electrons in this case
can be written as
\be
\hat{\G}^{R,A}(\k ,\, \epsilon) = \frac{1}{2}\sum_{s=\pm}
\frac{\hat{\sigma}_0 + s\lp\cos{\hat{\chi}_\k}\hat{\sigma}_1-
\sin{\hat{\chi}_\k}\hat{\sigma}_2\rp}{\epsilon - \xi^s_\k \pm i/2\tau}
\label{Green2}
\ee
where, $\cos{\hat{\chi}_\k}=\frac{\lambda \sin \theta+\beta \cos \theta
}{\lambda_{RD} }$
  and    $\sin{\hat{\chi}_\k}=\frac{\lambda \cos \theta +\beta 
\sin \theta}{\lambda_{RD} }$ .

We proceed to calculate spin-spin Hall conductivities in the same way as
in section III, 
but the calculation is much more cumbersome due to the angle dependent
spin-orbit coupling strength $\lambda_{RD}(\theta)$ in the dispersion.
We however find that one can avoid a lot of complication in 
self-consistent evaluation of the vertex corrected currents,
by concentrating on the 
largest order in $\Delta$. 
As long as the conductivities 
are non-zero in this order, the lower orders may be ignored. 
The vertex-corrected in-plane spin current components in the dc limit
are thus found to be
\bea
\hat{J}_1^x=\hat{J}_2^y &\approx& -
\frac{2\Delta}{B}\,\,\frac{\lambda B+\beta\lp\sqrt{A^2-B^2}-A
\rp}{\sqrt{A^2-B^2}-1}\hat{\sigma}_3 \\
\hat{J}_2^x=\hat{J}_1^y &\approx& 
-\frac{2\Delta}{B}\,\,\frac{\beta B+\lambda\lp\sqrt{A^2-B^2}-A
\rp}{\sqrt{A^2-B^2}-1}\hat{\sigma}_3
\eea
with $A=1+4\delta^2+4\delta_D^2$, $B=8\delta\delta_D$,
and $\delta_D = \beta k_F\tau$.

The $dc$ spin-spin Hall conductivities, defined in Eqs.~(\ref{spinperp}) and
(\ref{spinparal}) can be obtained now via retarded current-current correlation 
function (\ref{Pi}). We find
\bea
\sigma^\perp_{xy} &\approx& -\frac{\Delta}{8\pi}\lp \frac{
 A^2-B^2-A}{B^2 }
    \rp \frac{\sqrt{A^2-B^2}-A}{\sqrt{A^2-B^2}-1} \\
\sigma^\parallel_{xy} &\approx& -\frac{\Delta}{8\pi}\lp \frac{1}{B}
    \rp \frac{\sqrt{A^2-B^2}-A}{\sqrt{A^2-B^2}-1}
\eea

The fact that $\sigma^\parallel_{xy}$ is non-zero suggests that the spin 
polarization of the induced spin current in the transverse direction is 
not fully perpendicular to the applied spin current. It is, in fact, 
fully perpendicular when one of the couplings is absent. Therefore in 
realistic situations, where $\beta \ll \lambda$, the spin polarizations 
of the applied and induced spin currents are almost perpendicular to each 
other as $\sigma^\parallel_{xy}/\sigma^\perp_{xy} 
\sim \delta_D/[\delta (1+4\delta^2)] \ll 1 $ in this limit.



\end{document}